\documentclass[a4paper,11pt]{article}
\usepackage{pos}

\newcommand{\dd}{\text{d}}

\title{Contribution to the Proton Spin Puzzle from Helicity at small Bjorken $x$}

\author*[a,b]{Yossathorn Tawabutr}

\affiliation[a]{Department of Physics, University of Jyv\"askyl\"a,  \\
P.O. Box 35, 40014 University of Jyv\"askyl\"a, Finland}

\affiliation[b]{Helsinki Institute of Physics, \\
P.O. Box 64, 00014 University of Helsinki, Finland}

\emailAdd{yossathorn.j.tawabutr@jyu.fi}

\abstract{
The contribution to proton helicity from the spins of quarks and gluons at small Bjorken $x$ can be conveniently studied through the small-$x$ helicity evolution~\cite{Cougoulic:2022gbk}. Recently, it has been employed in a global analysis~\cite{Adamiak:2023yhz} that includes polarized inclusive and semi-inclusive deep-inelastic scattering measurements at small $x$. The analysis demonstrates an excellent agreement between the evolution and the data, but its predictions contain large uncertainties owing to the large number of free parameters in the initial conditions. The latest development~\cite{Dumitru:2024} aims to mitigate this issue by employing the valence quark model in the polarized proton target in order to replace most of the free parameters in the initial conditions by the physical information from the proton model.
}

\FullConference{31st International Workshop on Deep Inelastic Scattering (DIS2024)\\
 8–12 April 2024\\
Grenoble, France\\}

\begin{document}
\maketitle

\section{Background}

The proton spin puzzle is a longstanding problem in particle physics concerning the amount of spin and orbital angular momenta from quarks and gluons that contribute to the spin of the proton~\cite{Ji:2020ena,Jaffe:1989jz}. Following the Jaffe-Manohar decomposition, we can write the proton's spin as~\cite{Jaffe:1989jz}
\begin{align}\label{JM_decomp}
&\frac{1}{2} = S_q + S_G + L_q + L_G,
\end{align} 
where $S_q$ ($L_q$) and $S_G$ ($L_G$) are the spin (orbital) angular momenta of quarks and gluons inside the proton, respectively. The focus of this article lies on the first two terms, $S_q$ and $S_G$, in the helicity basis, specifically from the quarks and gluons at small Bjorken $x$. In particular, quark and gluon helicity inside the proton can be written in terms of the helicity-dependent parton distribution functions (hPDFs) such that
\begin{align}\label{hPDF}
S_q(Q^2) &= \frac{1}{2}\int\limits_0^1\dd x \, \Delta\Sigma(x,Q^2) \;\;\;\;\;\text{and}\;\;\;\;\; S_G(Q^2) = \int\limits_0^1\dd x \, \Delta G(x,Q^2) \, .
\end{align}
Here, $\Delta G(x,Q^2)$ is the gluon hPDF, while $\Delta\Sigma(x,Q^2)$ is the flavor singlet quark hPDF, which is the sum of hPDFs of all quark and antiquark flavors,
\begin{align}\label{S_hPDF}
\Delta \Sigma(x,Q^2) &= \sum_f\left[\Delta q_f(x,Q^2) + \Delta\bar{q}_f(x,Q^2) \right] .
\end{align}
The hPDFs are relatively well known at moderate values of $x$, but their results at small $x$ are lesser known due to the lack of experimental measurements at high center-of-mass energies. This limitation will be improved by the upcoming Electron-Ion Collider (EIC)~\cite{Accardi:2012qut,AbdulKhalek:2021gbh,Abir:2023fpo}. The existing knowledge of the total quark and gluon spins are such that~\cite{Aschenauer:2013woa,Aschenauer:2015eha}
\begin{subequations}\label{Sq_SG}
\begin{align}
S_q(Q^2=10\text{ GeV}^2) &\simeq \frac{1}{2}\int\limits_{0.001}^1\dd x \, \Delta\Sigma(x,Q^2=10\text{ GeV}^2) \in [0.15,0.20] \, , \\
S_G(Q^2=10\text{ GeV}^2) &\simeq \int\limits_{0.05}^1\dd x \, \Delta G(x,Q^2=10\text{ GeV}^2) \in [0.13,0.26] \, .
\end{align}
\end{subequations}
This implies that the sum of $S_q$ and $S_G$ cannot account entirely for the proton spin of $\frac{1}{2}$ according to Eq.~\eqref{JM_decomp}. The remaining contribution could come from the orbital angular momenta, $L_q$ and $L_G$, or the small-$x$ contributions that are missing from Eqs.~\eqref{Sq_SG}. This article summarizes recent developments~\cite{Cougoulic:2022gbk,Adamiak:2023okq,Adamiak:2023yhz,Dumitru:2024} aimed to quantify the latter.

Before we proceed, it is useful to consider the flavor \emph{non-singlet} combinations of the quark hPDFs, which in our research program are taken to be the difference between the quark and antiquark hPDFs for each flavor~\cite{Kovchegov:2016zex},
\begin{align}\label{NS_hPDF}
\Delta q^-_f(x,Q^2) &= \Delta q_f(x,Q^2) - \Delta\bar{q}_f(x,Q^2) \, .
\end{align}
These flavor non-singlet hPDFs are relevant in describing semi-inclusive processes.

\section{Parton Helicity at Small $x$}

In order to access parton helicity at small Bjorken $x$ inside the proton, it is convenient to modify the dipole formalism~\cite{Mueller:1989st,Nikolaev:1990ja} to incorporate helicity dependence into the study of deep-inelastic scattering (DIS) at high center-of-mass energy~\cite{Kovchegov:2015pbl,Kovchegov:2018znm,Cougoulic:2022gbk}. This introduces the \emph{sub-eikonal} corrections -- suppressed by a factor of the large light-cone momentum -- to the eikonal Wilson line that describes the color rotations taking place as a parton interacts with the target over a short light-cone time scale~\cite{Kovchegov:2012mbw}. These sub-eikonal, \emph{polarized} Wilson lines are formally defined in~\cite{Cougoulic:2022gbk}, and they induce the definition of \emph{polarized dipole amplitudes} of type 1, denoted by $Q_f(r_{\perp},zs)$ in the fundamental representation and $G(r_{\perp},zs)$ in the adjoint representation. As for the type-2 polarized dipole amplitude, the fundamental and adjoint representations yield the same dipole amplitude denoted by $G_2(r_{\perp},zs)$. Here, $r_{\perp}$ is the transverse dipole size and $zs$ is the center-of-mass energy of the interaction between the target and the polarized (anti)quark in the dipole.

In~\cite{Cougoulic:2022gbk}, the quark and gluon hPDFs have been shown to relate to polarized dipole amplitudes via the relations,\footnote{Recently, in~\cite{Borden:2024bxa}, the quark hPDF, $\Delta\Sigma(x,Q^2)$, is re-written in term of a new type of dipole amplitude, ${\widetilde Q}(r_{\perp},zs)$. The difference between the new and the current formalism does not show up until three loops in the anomalous dimension. However, the quantitative impact on global analysis remains to be seen in a future work.}
\begin{subequations}\label{hPDF_dip}
\begin{align}
\Delta\Sigma(x,Q^2) &= -\frac{N_c}{2\pi^3}\sum_f\int\limits_{\Lambda^2/s}^1\frac{\dd z}{z} \int\limits_{1/zs}^{\min\{1/zQ^2,1/\Lambda^2\}} \frac{\dd r^2_{\perp}}{r^2_{\perp}} \left[Q_f(r_{\perp},zs) + 2G_2(r_{\perp},zs)\right] , \\
\Delta G(x,Q^2) &= \frac{2N_c}{\alpha_s\pi^2} \left[\left(1+r^2_{\perp}\frac{\partial}{\partial r^2_{\perp}} \right) G_2\left(r_{\perp},zs=\frac{Q^2}{x}\right) \right]\Big|_{r^2_{\perp}=1/Q^2} \, .
\end{align}
\end{subequations}
This allows for the study of helicity at small Bjorken $x$ to be carried on in terms of the polarized dipole amplitudes and only perform the conversion at the end using Eqs.~\eqref{hPDF_dip}. In~\cite{Cougoulic:2022gbk}, a high-energy evolution equation has been derived to relate the polarized dipole amplitudes at high rapidity, corresponding to small $x$, to the \emph{initial conditions} of the dipole amplitudes at moderate $x$. Each step of this evolution resums $\alpha_s\ln^2(1/x)$, which is a factor that becomes large already at $x\lesssim 0.1$. It is worth emphasizing that this evolution evolves the dipole amplitudes from moderate to small Bjorken $x$, while keeping the resolution scale, $Q^2$, fixed. In the mean-field large-$N_c\& N_f$ limit~\cite{Veneziano:1976wm}, the evolution equation becomes a closed system of integral equations, involving the three polarized dipole amplitudes -- $Q_f(r_{\perp},zs)$, $G(r_{\perp},zs)$ and $G_2(r_{\perp},zs)$ -- together with their neighbor dipole amplitude counterparts -- $\bar{\Gamma}_f(r_{\perp},r'_{\perp},z's)$, $\Gamma(r_{\perp},r'_{\perp},z's)$ and $\Gamma_2(r_{\perp},r'_{\perp},z's)$, respectively -- which are the polarized dipole amplitudes with physical transverse size $r_{\perp}$ but lifetime $r'^2_{\perp}z'$, with $r'_{\perp}\leq r_{\perp}$.\footnote{See~\cite{Kovchegov:2015pbl,Kovchegov:2016zex} for more detailed introduction and discussion of the neighbor dipole amplitudes. Also, see~\cite{Tawabutr:2023lyd,Tawabutr:2023sei} for a more detailed summary of the small-$x$ helicity evolution equation.} Specifically, the complete evolution equation at large $N_c\& N_f$ is given explicitly in Eqs.~(155) of~\cite{Cougoulic:2022gbk}. 

Similarly, the flavor non-singlet hPDFs are relevant to helicity-dependent semi-inclusive DIS (SIDIS) processes. These hPDFs relate to the \emph{non-singlet} polarized dipole amplitudes, namely~\cite{Kovchegov:2016zex}
\begin{align}\label{hPDF_dip_NS}
\Delta q^-_f(x,Q^2) &= \frac{N_c}{2\pi^3} \int\limits_{\Lambda^2/s}^1\frac{\dd z}{z} \int\limits_{1/zs}^{\min\{1/zQ^2,1/\Lambda^2\}}\frac{\dd r^2_{\perp}}{r^2_{\perp}} \, G_f^{\text{NS}}(r_{\perp},zs) \, ,
\end{align}
where $G_f^{\text{NS}}$ is the flavor non-singlet polarized dipole amplitude. Its fundamental and adjoint representations are equivalent. At large $N_c\& N_f$, it obeys an integral evolution equation in $x$ that resums $\alpha_s\ln^2(1/x)$, akin to its flavor singlet counterparts.

\section{Global Analysis}

In order to compare the evolution equation from~\cite{Cougoulic:2022gbk} with experimental data and make realistic predictions about parton helicity at small $x$, a global analysis is performed in~\cite{Adamiak:2023yhz} based on the large-$N_c\& N_f$ evolution equations mentioned above. In that work, the polarized dipole amplitudes at moderate $x=0.1$ are taken to be different linear combinations of transverse and longitudinal logarithms,
\begin{subequations}\label{IC}
\begin{align}
Q_f^{(0)}(r_{\perp},zs) &= a_f\,\ln\frac{zs}{\Lambda^2} + b_f\,\ln\frac{1}{r^2_{\perp}\Lambda^2} + c_f \, , \\
G^{(0)}(r_{\perp},zs) &= a\,\ln\frac{zs}{\Lambda^2} + b\,\ln\frac{1}{r^2_{\perp}\Lambda^2} + c \, , \\
G_2^{(0)}(r_{\perp},zs) &= a_2\,\ln\frac{zs}{\Lambda^2} + b_2\,\ln\frac{1}{r^2_{\perp}\Lambda^2} + c_2 \, , \\
G_f^{\text{NS}(0)}(r_{\perp},zs) &= a_f^{\text{NS}}\,\ln\frac{zs}{\Lambda^2} + b_f^{\text{NS}}\,\ln\frac{1}{r^2_{\perp}\Lambda^2} + c_f^{\text{NS}}\, ,
\end{align}
\end{subequations}
where $\Lambda$ is an infrared scale. Here, $a$'s, $b$'s and $c$'s make up the 24 free parameters of the global analysis. These parameters are fixed by matching the evolved dipole amplitudes with all available polarized DIS and SIDIS measurements at $x\leq 0.1$. The inclusion of the polarized SIDIS data required for the flavor non-singlet dipole amplitudes to be considered as well as its flavor singlet counterparts. In total, 226 data points are available for different observables, including $A_1$ and $A_{\parallel}$ for polarized DIS together with $A_1^h$ for polarized SIDIS. Targets include proton, deuteron and helium-3, while for SIDIS charged pions, charged kaons and unidentified charged hadron production data are considered. The analysis are performed using the Monte Carlo Bayesian analysis under the Jefferson Laboratory Angular Momentum (JAM) framework~\cite{Adamiak:2023yhz}.

Overall, the small-$x$ helicity evolution equations at large $N_c\&N_f$~\cite{Cougoulic:2022gbk,Kovchegov:2016zex} together with the initial conditions given in Eqs.~\eqref{IC} lead to an excellent agreement with experimental data, resulting in $\chi^2=1.03$ per number of data points~\cite{Adamiak:2023yhz}. The main prediction of the global analysis includes the total spins of quarks and gluons inside the proton down to $x = 10^{-5}$, which is at least two orders of magnitude smaller than the minimum Bjorken $x$ from the RHIC spin program~\cite{Aschenauer:2013woa,Aschenauer:2015eha}, c.f. Eqs.~\eqref{Sq_SG}. With the free parameters in Eqs.~\eqref{IC} fixed by the data, the evolution equations yield the dipole amplitudes at small $x$, which in turn yield the quark and gluon hPDFs via Eqs.~\eqref{hPDF_dip}. Together with uncertainty propagated through the calculation, we obtain
\begin{align}\label{spin_num}
&\int\limits_{10^{-5}}^{0.1}\dd x \left[\frac{1}{2}\Delta\Sigma(x,Q^2=10\text{ GeV}^2) + \Delta G(x,Q^2=10\text{ GeV}^2)\right] = -0.64 \pm 0.60 ,
\end{align}
slightly favoring the scenario with net negative helicity coming from partons at small $x$. Note that the left-hand side of Eq.~\eqref{spin_num} with upper limit $x=1$ would yield a good approximation for $S_q+S_G$, c.f. Eqs.~\eqref{hPDF}. However, the uncertainty remains relatively large, allowing for limited qualitative interpretation of the results in the context of the proton spin puzzle.

The large uncertainty in the predictions from global analysis~\cite{Adamiak:2023yhz} results from the significant flexibility in the model for the initial conditions~\eqref{IC}, containing 24 free parameters. In order to mitigate the issue, a recent moderate-$x$ calculation has been performed in Ref.~\cite{Dumitru:2024} with the goal of replacing the majority of the free parameters by physical information of the proton target. In particular, the light-front valence quark model~\cite{Dumitru:2018vpr} has been employed, including the corrections coming from a gluon emission~\cite{Dumitru:2020gla}. This is the first time such a model is employed within the polarized dipole framework~\cite{Dumitru:2024}. As a result, the physical input from the valence quark model is capable of fixing the coefficient of the transverse logarithmic term, $\ln(1/r^2_{\perp}\Lambda^2)$, in all the polarized dipole amplitudes, while drawing relations among coefficients of other terms. At the end, the resulting initial conditions contain 3--9 free parameters, depending on the flexibility of the model~\cite{Dumitru:2024}, which is a significant reduction from the initial conditions~\eqref{IC} employed in~\cite{Adamiak:2023yhz}.\footnote{It is straightforward to generalize the calculation in Ref.~\cite{Dumitru:2024} to include the new type of dipole amplitude defined in Ref.~\cite{Borden:2024bxa}, which corresponds to sub-eikonal quark exchanges with an either light-cone forward or backward staple.} Furthermore, all the parameters fixed by the valence quark model have their values within the 95\% confidence interval of their respective values fitted by the global analysis~\cite{Adamiak:2023yhz}. This is a promising development that could reduce the uncertainty in the predictions of future global analyses for helicity at small $x$.

\section{Conclusion and Future Work}

In this article, we summarize the current development of our understanding of parton helicity at small $x$ based on the small-$x$ helicity evolution~\cite{Kovchegov:2016zex,Cougoulic:2022gbk}. The recent global analysis~\cite{Adamiak:2023yhz} produces a great agreement between the evolution formalism and experimental measurements, but its physical predictions still contain large uncertainties, which are likely due to the large number of free parameters in the initial conditions. 

To tackle this challenge, two main directions have emerged. First, Ref.~\cite{Dumitru:2024} calculates a new set of initial conditions for the polarized dipole amplitudes with significantly fewer free parameters, while replacing the rest of the parameters by the physical input from the light-front valence quark model of the proton target. In light of this development, a future global analysis based on the new initial conditions from~\cite{Dumitru:2024} and the small-$x$ helicity evolution~\cite{Kovchegov:2016zex,Cougoulic:2022gbk} is expected to yield physical predictions with significantly reduced uncertainties. 

Another promising development is to include longitudinal spin asymmetries for particle productions in proton-proton collisions into the global analysis, providing additional data sets to further constraint the parameters. At the current stage, only the pure-gluon channel is available~\cite{Kovchegov:2024aus}, while the inclusion of quarks remains a work in progress.

Given the excellent agreement in Ref.~\cite{Adamiak:2023yhz} between the evolution equation and the experimental data, together with the two directions of development aimed at improving the precision of the helicity global analysis, it is optimistic that a new global analysis in the near future will provide meaningful and precise physical predictions that allow for a better understanding into the proton spin puzzle. Longer term in the future, our understanding will receive further improvements with the addition of the single-logarithmic corrections to our evolution equation, c.f.~\cite{Kovchegov:2021lvz,Tawabutr:2021xrr}, together with the upcoming EIC measurements of polarized DIS and SIDIS at small Bjorken $x$~\cite{Accardi:2012qut,AbdulKhalek:2021gbh,Abir:2023fpo}.

\providecommand{\href}[2]{#2}\begingroup\raggedright

\end{document}